# The relationship between the compressional and shear strengths of poroelastic colloidal gels

Richard Buscall

MSACT Consulting, Exeter, U.K.
Email: r.buscall@physics.org

*Strong particulate gels are widely believed to behave poroelastically in compression, e.g. in sedimentation, even though they consolidate irreversibly because of the stickiness of the particles. Particulate gels are usually adhesive as well as cohesive and so wall effects are to be expected in general (Michaels & Bolger 1962 [3]). These are rarely manifest on process engineering scales, although they can be important in the laboratory and with formulated products in small containers. When the assumption of non-linear poroelasticity is combined with the idea that adhesive failure (yield) is brittle, relatively speaking, a prescription emerges for the ratio of shear to compressive strength S and how this varies with density. S is predicted to be of order unity at the gel-point and then to increase rapidly thereafter.*
*The predictions are consistent with the experimental data available for both dilute and concentrated gels, although more data would be welcome. A critique of a recent paper by Condre et al. 2007 [30] on wall effects in very dilute gels is given in which it is argued, amongst other things, that it is not necessary to invoke granular in order to account for their results, simple adhesion suffices.*

## Introduction

Above some critical concentration, the gel-point, the aggregation of sticky colloidal particles results in the formation of a ramified particulate gel [1]. Such gels may or may not be stable in respect of sedimentation (or creaming), depending *inter alia* upon a balance between the strength of the gel and its self-weight [2]. Because the particles are sticky, both the compressional strength and the (adhesive) shear strength are important [3], since the gel is likely to adhere to the walls any container, albeit to a degree that might depend upon the material nature of the walls. The stability is known to depend also upon the strength of the interparticle attraction, since, if this is sufficiently large, particulate gels behave poroelastically [4,5], whereas weak gels can show delayed collapse [6-11] and creeping sedimentation [5,12,13]; both manifestations of restructuring as a result of fluctuations. This communication will be concerned with strong poroelastic gels only.

**Theory**

The phenomenological theory of the sedimentation of strong particulate gels is now well-established, even though this might well not be obvious from a superficial reading of what is a rather large and fragmented literature [e.g. 14-29, loc. cit. ]. A thorough critical review of the theory would divert from the purpose of this modest paper, which is to examine the origins of the relationship between shear and compressive strength, and so here, only the salient features will be described, following [15].

The pressure at some point $z,r$ in a sedimenting gel is given by,

$$p(z,r) - p_0(z) = p_s(z,r) + p_l(z,r) - p_0(z) \leq \Delta\rho g \varphi (H - z) \qquad (1$$

where $p_0$ is the hydrostatic pressure in the absence of the particles and the subscripts $s$ and $l$, denote the particulate and liquid phases, $\Delta\rho$ is the difference in density between solid and liquid, $\varphi$ is the local volume-fraction of solid and $H$ is the height of the gel. If the liquid is taken to be incompressible then $p_0(z)$ can be set to zero, whence $p, p_s,$ and $p_l$ are the excess pressures arising from the density mismatch between solid and liquid phases. The inequality in eqn 1 arises because in general the gel will be supported in part by the walls. If so, there will be a shear stress at the walls, $\tau_w$. A momentum balance on the particulate phase then gives [3,14],

$$0 = \Delta\rho g \varphi + \frac{\partial p_s}{\partial z} - \Gamma u - \frac{4\tau_w}{D} \qquad (2$$

where, $u$ is the local sedimentation velocity of the particulate phase, $\varphi = \varphi(z)$ is the solid phase volume fraction at a level $z$ and $\Gamma = \Gamma(\varphi)$ is the friction coefficient per unit volume. The latter can variously be written in terms of, either, the Stokes' drag and the hindered settling function [e.g. 15, 25], or, the Darcy's law permeability [e.g. 24], or, the hydraulic conductivity [e.g. 23], as discussed elsewhere [26, 27] and is summarised in appendix 2. Notice that eqn 2 is one-dimensional, even though the shear stress $\tau_{rz}$ varies radially, it makes a constant contribution to the pressure gradient, i.e. $\frac{\tau_{rz}(r,z)}{r} = \frac{\tau_w}{D}$. For the present purposes eqn 2 will suffice since it is not proposed to consider the internal shear strains. The assumption made implicitly here is that the critical strain for yield at the wall is small, a point which will be amplified and justified below.

It is standard to decompose the solid phase pressure as follows [15],

$$\frac{\partial p_s}{\partial z} = \frac{dp_s}{d\varphi}\frac{\partial \varphi}{\partial z} \equiv K\frac{d\ln\varphi}{dz} \qquad (3$$

where, now, $p_s$ is to be identified with the equilibrium compressional strength, $p_y$, and where $K$ is the so-called compressional modulus [4, 15]. To be more precise, it is the bulk modulus of the particulate phase: not the unaxial compressional modulus, as is sometimes claimed, for reasons that are outlined appendix 1. $p_y$ is sometimes called the compressional yield stress in recognition of the irreversible nature of consolidation, it is important to emphasize though that there is no elastic strain limit; the network becomes stronger as it densifies and irreversibility is purely a consequence of the stickiness of the particles. Kim et al. [24] have coined the term "ratchet poroelasticity" in order to capture the idea of irreversible poroelastic consolidation. It is important to realise though that the "compressional yield stress" [15, 19, 20,21] and "ratchet poroelastic" [24, 28, 29] formulations are identical; trivially so, in the light of eqn. 3. Thus, for example, if the permeability coefficient and modulus in Kim et al. [24,29] and are replaced by the Stokes' drag plus hindered settling function and the compressional yield stress (via eqn 3) respectively, the equations of Buscall and White [15] are recovered. At least they are, if gel possesses strength as well as elasticity from the outset, i.e., $K(\varphi_0) > 0$ and $p_y(\varphi_0) > 0$, which it should provided that is that $\varphi_0$ is above the gel point $\varphi_g$. This is mentioned because in some work there has been a tendency in [e.g. 24, 28-30] to assume that $K(\varphi_0) > 0$ but $p_y(\varphi_0) = 0$ for the starting gel, which is likely only to be valid very near $\varphi_g$. This point and its implications will be returned to in appendix 3 (see also eqns 11 and 12 below). The various representations of the drag function are shown in appendix 2 following [26] *loc cit*. Further details of the general 1-d poroelastic sedimentationmodel can be found variously [14-29 and 51-57], although there is much duplication and redundancy to be found. A good introductory review was provided earlier by Landman and White [58] and an updated overview of sedimentation is in hand. Note though, that the basic is a unifying model of 1-d solid liquid separation operations more generally [23, 27, 45, 58] and the materials functions are portable.

The sedimentation of strong gels then is controlled by three material functions then, the drag function $\Gamma(\varphi)$, a wall friction $\tau_w(\varphi)$ and a compressional strength function; either $K(\varphi)$, or, its integral over the volumetric Hencky strain, $p_y(\varphi)$, according to taste. It is important to emphasise that all three material parameters are functions and not constants and strong functions of concentration, to boot. Because of this, poroelastic sedimentation theory should not be linearized except for didactic purposes, even where the experimental parameter space is very small [28]. Even then, care has to be taken because the linearized results can be misleading if used out of context [30], as will be demonstrated in appendix 3. Asymptotic and approximate solutions to the sedimentation eqns and the background to and details of numerical methods can be found in various works [15,19,20,28, 51-58]. For example, the initial boundary sedimentation rate in absence of wall effects, i.e., the limit of $4\tau_w H_0 / p_y D \ll 1$ is given by [15],

$$\left. \frac{-dH}{dt} \right|_{t=0} = \min \left\{ \left| \frac{\Delta \rho g \varphi_0}{\Gamma(\varphi_0)} \left[ 1 - \frac{p_y(\varphi_0)}{\Delta \rho g \varphi_0 H_0} \right] \right|, 0 \right\} \qquad (4$$

This result is readily generalised to include wall friction in the limit of negligible shear strain to give,

$$\left.\frac{-dH}{dt}\right|_{t=0} = \min\left\{\left|\frac{\Delta\rho g\varphi_0}{\Gamma(\varphi_0)}\left[1-\frac{p_y(\varphi_0)}{\Delta\rho g\varphi_0 H_0}-\frac{\tau_w(\varphi_0)}{\Delta\rho g\varphi_0 D}\right]\right|, 0\right\} \quad (4a$$

These two equations, 4a and 4b, are analogous to the Bingham eqn for the flow of a yield stress material, in effect, since, in both cases, the total stress is decomposed into a linear viscous part and an elastic part. Good approximations are available also for the long-time kinetics, i.e. the approach to equilibrium, see Usher et al. [20] for further details, and various asymptotic solutions are discussed in Howells et al. [19] and Buscall and White [15].

The condition for sedimention equilibrium is given by setting the velocity $u$ to zero everywhere in eqn 2 to give,

$$0 = \Delta\rho g\varphi + \frac{dp_y}{d\varphi}\frac{\partial\varphi}{\partial z} - \frac{4\tau_w}{D}, \quad (5$$

noting also that, by conservation of mass,

$$0 = \frac{1}{\varphi_0 H_0}\int_0^{H_\infty}\varphi(z)dz - 1, \quad (6$$

where $H_\infty$ is the equilibrium sediment height, which can then be obtained by integration, given correlations for the two material coefficients. One such correlation is,

$$p_y = k\left(\frac{\varphi}{\varphi_g}\right)^n - 1; \quad K = \frac{k}{n}\left(\frac{\varphi}{\varphi_g}\right)^n = (p_y+1)/n \quad (7$$

This captures the rapid development of strength just above the gel-point and the power-law behaviour (n ≈ 4 typically) seen at somewhat higher concentrations, although not the divergence expected at the glass transition (e.g. 0.59 for monodisperse hard-spheres). In the absence of wall friction Eqns 5-7 can be integrated to give [20],

$$\frac{H_\infty}{H_0} = \varepsilon\left\{1 + \frac{C^n}{(n-1)(C^n-1)}\left[\left(1 + \frac{n(1-\varepsilon)(C^n-1)}{\varepsilon C^n}\right)^q - 1\right]\right\}; \quad n > 1 \qquad (8a)$$

$$= \varepsilon\left\{1 + \frac{C}{(C-1)}\left[\ln\left(1 + \frac{1}{\varepsilon(C-1)}\right) - \ln(C)\right]\right\}; \quad n = 1 \qquad (8b)$$

where the exponent in eqn 8a is $q = (n-1)/n$, not $n/(n-1)$ as shown in [20] and where $\varepsilon = \frac{p_y}{\Delta\rho g \varphi_0 H_0}$ is the scaled compressional strength. In a limited sense, eqn 8b for n=1 represents a linearization of 8a, inasmuch that the strength and the modulus increase in proportion to the volume-fraction (eqn. 7) and the gel has strength as well as rigidity at $\varphi_0$ (eqn 7 likewise), whereas the the linearisation used by Manley et al. [28] and Condre et al. [30], $K$ = constant, $p_y = K(\varphi - \varphi_0)/\varphi_0$, is very different in several respects. For example, whereas eqn 8b is exact, the equations of Manley et al. [28] and Condre et al. [30], are only valid in the limit, either, of negligible compression, i.e $\frac{H_\infty}{H_0} \simeq 1$, or, of dominating wall-friction and they can be very misleading if used otherwise, even though data can sometimes be fitted to them for one reason or another[28]. These matters will be taken up in appendix 3.

Since the gel will normally be adhesive, it is reasonable to equate the wall stress $\tau_w$ with the yield stress of the gel in shear, as was done by Michaels and Bolger [3], although it needs to be recognised that the *adhesive* yield stress $\tau_w$ may differ from the material or "true" yield stress $\tau_y$. This difference is exploited in the vane method of determining yield stresses [31] whereby a large gap between vane and cylinder is used avoid premature yield at the wall of the cylinder. On the other hand, a small gap can be used deliberately to determine $\tau_w$ also [32], supposing, that is, that $\tau_w < \tau_y$, which appears usually to be the case [32-35]. $\tau_w$ and $\tau_y$ are expected to be the same order of magnitude though and so here it will be assumed that $\tau_w \approx \tau_y$. This assumption will be made in order to extend the range of experimental data available, since in some cases one has been measured and in some cases the other (and in yet others, which is unclear). The strain at yield will be of concern later. Whereas it appears to be the case that particulate gels yield over a range of stress and strain [32, 36], indicating that the critical condition is not simply a critical stress or total strain condition, it is nevertheless always possible to identify a maximum level of strain above which flow always occurs. The critical shear strain is usually found to be small [32, 36-38], typically between 0.0001 and 0.01, although larger values have been reported for very dilute dispersions of very small particles, i.e. true nanoparticles [30]. The critical strains for adhesive failure are found to be very similar in magnitude to those for cohesive failure [32]. Here, $\gamma_c$ will be used to denote the critical shear strain, without making a distinction between adhesive and cohesive failure, whereas $\varepsilon_r$ will be used to denote any recoverable volumetric strain.

The processes of shear yield and irreversible compression are very different then. For the presence purposes it will suffice to ignore the possibility that the precise value of the shear yield stress might be time-dependent due to restructuring [36] and assume that the yield stress is related to the critical strain and shear modulus $G$ by,

$$\tau_y(\varphi) = \int_0^{\gamma_c} G(\varphi,x) \mathrm{d}x \qquad (9$$

Or, if we assume linearity up to the yield point, for the sake of argument,

$$\tau_y(\varphi) \leq G(\varphi,0)\gamma_c. \qquad (10$$

In compression on the other hand, the structure becomes stronger as it densifies and both the strength and stored elastic energy increase with φ. It may not even be necessary to break many if any bonds even transiently; the structure may just crumple like a ball of paper being squeezed. Be that as it may, the system strain hardens strongly and without any relaxation in the strong aggregation limit. The densification is irreversible simply because the particles are sticky.

We then write the total volumetric strain for large strains as,

$$\varepsilon_{total} = \int_{\varphi_g}^{\varphi} d\ln x = \varepsilon_r(\varphi_0) + \int_{\varphi_0}^{\varphi} d\ln x \tag{11}$$

and the pressure at some volume fraction φ just at the "yield" point is,

$$p_y(\varphi) = \int_{\varphi_g}^{\varphi} K(x) \, d\ln x = \varepsilon_r(\varphi_0) K_u(\varphi_0) + \int_{\varphi_0}^{\varphi} K(x) \, d\ln x \tag{12}$$

where $K_u(\varphi)$ is the bulk modulus that would be seen upon unloading, were tthat to be possible. Note that if the recoverable strain is small, then $K_u(\varphi) > K(\varphi)$. $P_y$ (or $K$) can variously be measured by either batch sedimentation or centrifugation [4,15] or pressure filtration [39]. In the case of the latter two methods it would be possible to measure the recoverable volumetric strain in principle. In practice it is normally found that $\varepsilon_r$ is not resolvable, implying that it is small. It should realised that were $\varepsilon_r$, say, to be of the same order as $\gamma_c$, say, then it would not be resovable upon unloading, since whereas strains as small as 10ppm can be resolved by modern shear rheometers, resolving a volume change of even 0.01 is difficult in batch sedimentation or centrifugation.

From eqn 10 and 12 the ratio of the compressive strength to the shear strength is given by,

$$S = \frac{\tau_y}{p_y} = \frac{\int_0^{\gamma_c} G(\varphi, x) dx}{\int_{\varphi_g}^{\varphi} K(x) \, d\ln x} \leq \frac{G(\varphi, 0)\gamma_c}{\int_{\varphi_g}^{\varphi} K(x) \, d\ln x} \tag{13}$$

It is fairly evident from eqn 13 that if the strains $\gamma_c$ and $\varepsilon_r$ are small, but comparable, then S would decrease from $S_g \sim G/K$ near the gel-point to ever smaller values as the volume-fraction is increased. The next step is to relate $K$ to $G$. For isotropic elastic bodies, the bulk modulus *K* is related the shear modulus *G* by Poisson's ratio ν, according to,

$$K = \zeta G; \quad \zeta = \frac{2}{3}\left(\frac{1-\nu}{1-2\nu}\right); \quad \text{with } -1 \leq \nu \leq \tfrac{1}{2} \tag{14}$$

which gives an absolute lower limit for the value of *K/G* of 4/9, if auxetic behaviour is admitted, and a more realistic lower limit of 2/3 (cf. cork) otherwise. As for the upper limit, one estimate can be made by supposing that the network is held together by central forces, which might be thought a reasonable assumption for a network comprising colloidal particles, for moderate binding forces at least. For networks bound by central forces *K/G* is constrained to be 5/3 by Cauchy's relationships. So, to summarise, we might expect,

$$\frac{2}{3} \leq \frac{K}{G} \leq \frac{5}{3} \tag{15}$$

which in turn leads to the expectation that,

$$G(\varphi) \approx K(\varphi) \equiv \frac{dP_y(\varphi)}{d\ln\varphi} \tag{16}$$

This is what has been found very approximately [37, 40- 41,], as will be seen from the graphs in figs. 1 and 3 to 5, to be presented in more detail in the next section. Note that it is G′(ω), at a single frequency ω, which will plotted, but that the measurements in figs. 3-5 were made by a wave propagation method whereby a signal is only detected if the relaxation rate is negligible.

As an aside, it can be argued that the existence a correlation of the form of eqn 16 is important for several reasons: it gives confidence in our assumptions about the nature of the constitutive rheology underpinning consolidation theory, it reminds us that there is more to suspension rheology than just shear rheology, and it provides a way of estimating the compressive strength in the absence of direct measurements. The latter could be useful, in the context of weak gels for example, where creep prevents measurements of sedimentation equilibrium, except at high volume fraction. The latter show more complex behaviour (creep and delayed collapse [6-13]) than do strong gels, as a consequence of fluctuations, but since the effect of fluctuations is superimposed on top of what would be happening otherwise, estimates of the notional network strength form a benchmark, arguably. Whence it would be useful to know what value the strength have, either instantaneously, or, in the absence of fluctuations. Thus, because $G'(\omega)$ can be measured rapidly and at a frequency chosen to exceed the fluctuation rate, eqn 16 allows one to estimate what strength the gel would have and thus how stable it would be if the fluctuations were frozen and, specifically, to ask what the stress distribution between particulate and fluid phases would be under such circumstances.

Eqns 13 and 15 combine to give,

$$S = \frac{\tau_y}{p_y} = \frac{\int_0^{\gamma_c} G(\varphi,x)\mathrm{d}x}{\int_{\varphi_g}^{\varphi} K(x)\,\mathrm{d}\ln x} \leq \frac{5G(\varphi,0)\gamma_c}{3\int_{\varphi_g}^{\varphi} G(x,0)\,\mathrm{d}\ln x} \tag{17}$$

For example, if the moduli have the form given in eqn 7, then eqn 17 becomes,

$$S = \frac{\tau_y}{p_y} \leq \frac{5n\gamma_c}{3\left(1-\left[\frac{\varphi_g}{\varphi}\right]^n\right)}; \quad \varphi > \varphi_g \tag{18}$$

Away from the gel-point then, $S$ is predicted to be of order $n\gamma_c$ when the moduli show power-law behaviour at intermediate volume-fractions. What is to expected very near the gel-point is less certain since the details then become important, $S \sim 1$ would however seem likely.

**Comparison with experiment**

Whereas there have been a very large number of measurements of the functions $\tau_w(\varphi)$, $G(\varphi)$ and $p_y(\varphi)$ (or $K(\varphi)$) separately, there are relatively few sets of the three to be found in the literature. What there is, or, to be more precise, what the author could find at the time of writing, is shown in figs. 1-5. Figs. 1 and 2 show three sets of data by different workers for very similar colloidal aluminas close to their iso-electric points. The data of Channell and Zukoski [38] show $S$ to decrease from ca. 0.05 to 0.01 over the concentration range, whereas the Melbourne workers [42-45] see 0.2 down to ca. 0.01. Channell and Zukoski [38] also found the critical shear strain to be of order 0.01 and so it can be seen that the $S$ values are of order the critical shear strain, as expected. Fig. 3 shows similar data for coagulated polystyrene latex (PSL) [37, 40-41]. This shows much smaller $S$ values ranging from ca. 0.01 down to ca. 0.001. This would seem to be consistent with the observation that the critical shear strains for PSL [37, 40-41] are much smaller than those seen for alumina [38]. In the cases of nano-silica [41] and attapulgite clay [4, 15, 37] the shear strength was not measured, these plots showing the comparison between $G(\varphi)$, $p_y(\varphi)$ and $K(\varphi)$) only. Taking these data together with that in figs. 1 and 3, it can be seen that whereas these three functions are always similar in form and of the same order of magnitude, $G(\varphi)$ correlates better with $p_y(\varphi)$ than $K(\varphi)$ in two cases, and better with the latter in the other two.

Overall the data would appear to be reasonably consistent with expectation overall since they show, first, small $S$ values that decrease with increasing volume-fraction, second a good correlation between $G$ and $K$ in terms of concentration-dependence, and, third, a reasonable correspondence in terms of order of magnitude. The comparison is thus encouraging, so far as it goes. It is possible to make numerical predictions from, e.g. eqns 13 and 15, of course, given a value for the poisson's ratio etc. and this is done in fig. 6 for $n=4$ and $v=¼$ (i.e. $K/G=1$) for two typical (i.e. plausible) values of the critical shear strain.

More recently Manley et al. [28] and Condre et al. [30] have published sedimentation data coming from a different part of the parameter space, viz. for particles even smaller than those in fig. 5 and very low volume-fractions. Condre et al. [30] found that wall effects dominated in their case, implying values of $S$ of order unity, whereas Manley et al. [30], working with a very similar silica system at a somewhat higher volume-fraction found the compressive stress to play a role. Again, these observations would seem to be in accord with expectation, so far as they, go, especially when it is realised that these dilute nanoparticle gels show larger critical shear strains than is usual for colloidal systems (of order 0.1). The results of Condre et al. will be considered further in appendix 3 where certain errors in [30] will be addressed.

Finally, it should be mentioned that Channell and Zukoski [38], following, Meeten [46], have interpreted $S$ in terms of a "plastic" Poisson's ratio, a concept coming from soil mechanics. This approach is not however predictive in itself, one has to suppose that there is a critical strain associated with compressional "yield" and to say what it is, relative to the shear strain. The more physical appealing assumption of irreversible poroelasticity avoids this problem, since then the "critical" volumetric strain is the total strain. Furthermore, the simple poroelastic model elaborated here can be tested experimentally in considerable detail, given an appropriate data set. The latter would comprise the set $\tau_y(\varphi)$, $G(\varphi)$ and $p_y(\varphi)$ (or $K(\varphi)$) plus $\gamma_c$ measured over a wide range of volume-fraction from very close to the gel-point upwards. Such a set of data would give eqn 17 nowhere much to hide, except perhaps that the central-force upper limit on Poisson's ratio might need to be relaxed for strong cohesive forces, conceivably.

**Conclusions**

When the assumption of poroelasticity is combined with the idea that adhesive failure (yield) is always brittle, relatively speaking, then in is possible to come up with a prescription for the ratio of wall to compressive strength S and how this varies with density. S predicted to be of order unity at the gel-point and then to increase rapidly thereafter.

Overall the data would appear to be reasonably consistent with expectation overall since they show, firstly, small $S$ values that decrease with increasing volume-fraction, second, a good correlation between $G$ and $K$ in terms of concentration-dependence, and, third, a reasonable correspondence in terms of order of magnitude. The comparison is thus encouraging, so far as it goes.

A critique of a recent paper by Condre et al. 2007 [30] on wall effects in very tenuous gels is given in appendix 3, together with some corrected results and a discussion of the perils of linearising the equations of poroelastic sedimentation. It is argued, *inter*

*alia*, that it is not necessary to invoke granular friction in order to account for their results, simple adhesion suffices.

**Acknowledgements**  RB was self-supported (like a strong colloidal gel). The provision of electronic library facilities by the University of Leeds and a physical library by University of Exeter is gratefully acknowledged.

**Appendix 1 The nature of confined uniaxial loadings**.

There has been a tendency to assume or suppose [e.g. 15, 30] that because gravity acts axially in a sample confined in a vertical right annulus and because the radial displacements are necessarily zero at the inner surface of the annulus, the radial strain and stress must be zero also, i.e. that the loading is uniaxial. Whereas if the gel is elastic, from the equations of elastic stress equilibrium [47], this cannot be so unless the gel happened to be totally anisotropic. For isotropic gels the loading is hydrostatic necessarilyand $K$ is the bulk modulus, not the uniaxial compressional modulus $E_c$, say. There would seem to be no reason to assume other than isotropy without good empirical cause. Note that the distinction between $K$ and $E_c$ is unimportant for most intents and purposes since $K$ and $p_y$ are defined operationally in detailed models of 1-d poroelasticity [e.g. 15,19,20,27] and treated as quantities to be measured from sedimentation, centrigugation or pressure filtration experiments, i.e. the usage is self-consistent. Was, however the aim, say, to predict $K$ or $p_y$ from, say, particle-level models or simulations, or to related them to, say, $G$ as has been attempted here, then the distinction becomes important.

**Appendix 2 The drag function $\Gamma(\varphi)$.**

The drag coefficient $R_f(\varphi)$ can be written as [21, 35],

$$\Gamma(\varphi) = \frac{\varphi}{1-\varphi} R_f(\varphi) \qquad (A.2.1$$

$$R_f(\varphi) = \frac{(1-\varphi)\Delta\rho g}{u_{St} H(\varphi)} \qquad (A2.2a$$

where $u_{St}$ is the Stokes' law settling velocity of a single particle and $H(\varphi)$ is the hindered settling function, given e.g. by the Richardson Zaki correlation [26,48]. Eqn A2.2a assumes that the network is built of primary particles, which may not be the case and often is not. It might be written more generally perhaps, as,

$$R_f(\varphi) = \frac{(1-\varphi)\Delta\rho g}{u_{St} H(\varphi)} \frac{R^2}{R_C^2} \qquad (A2.2b$$

where the possibility that the network might be built of dense clusters of size of mean areal size $R_C$ has been included. Eqns A2.1 and A2.2 are one way to account for the drag in a flocculated system, but it is not the only way: thus some have preferred to use either the Darcy's law permeability coefficient $\kappa(\varphi)$ [e.g. 24, 28, 29], or the hydraulic conductivity $\Lambda(\varphi)$ [e.g. 23]. These are related to $R_f(\varphi)$ by [26, 27],

$$\kappa(\varphi) = \frac{\mu \Lambda(\varphi)}{\rho_f g} = \frac{(1-\varphi)\mu}{\varphi R_f(\varphi)} = \frac{u_{St}\mu H(\varphi)}{\Delta \rho g \varphi} \frac{R_C^2}{R^2} \qquad \text{(A2.3)}$$

Like the gel strength, drag function is normally regarded as a material property to be measured [27, loc. cit.] using methods developed by the group at Melbourne [49,50], although the sometimes correlations like Kozeny-Carman or Richardson-Zaki are used to represent or model the experimental drag function, or to interpolate where, say, gravity batch-settling data and pressure-filtration data do not overlap. The KC and RZ appear to work rather well in practice unless the particles happen to be coated with slime-layers as can be the case in water and waste-water treatment [27, *loc.cit.*].

**Appendix 3 Comments on linearisation [24] and corrections to Condre et al. [30].**

There is a temptation perhaps to linearize the poroelastic sedimentation model in order to obtain simple analytical results for the purposes of understanding, scaling etc. This has to be done with caution though and the results used with care, not least, because the linearized results tend to take on a life of their own, leading to the risk of their being used inappropriately. Because the self-weight is a maximum at the bottom (or, top, in the case of a creaming system), the volumetric Hencky strain at the bottom of a sediment can be quite large even for small degrees of boundary sedimentation, $\Delta = H_o - H(t)$. It is thus rarely, if ever, valid to compare linearized results with experimental data, even though this has been done [28,30] and with some success apparently, albeit for a limited experimental parameter space. That it appears to work at all is surprising but there are reasons for this as will now be shown using sedimentation equilibrium by way of example. Two steps are involved in linearizing, the first is to take $K(\varphi)$ to a constant $K$. This alone is severely limiting approximation, given that $K(\varphi)$ usually varies at least as fast as the third-plus power of $\varphi$ (cf. figs 1-5). The second step is to expand the Hencky strain, or, to put it another way, to replace it by the "engineering strain". Thus, summarising both steps, linearization involves,

$$p_y = \int_{\varphi_g}^{\varphi} K(x) d\ln x \;\Rightarrow\; K \ln\left(\frac{\varphi}{\varphi_g}\right) \;\Rightarrow\; K \ln\left(\frac{\varphi}{\varphi_0}\right) \;\Rightarrow\; \frac{(\varphi - \varphi_0)}{\varphi_0} K \qquad \text{(A3.1)}$$

where the intermediate steps have been included to emphasise that the final approximation on the RHS involves either assuming that $\varphi_0 \approx \varphi_g$, either explicitly or implicitly, or, that the initial, uniform gel has a modulus but no compressive strength for some reason. It should already be fairly clearly that eqn A3.1 is dangerous potentially, since, if it is used for other than very small maximum strains (which means negligible $\Delta = H_o - H(t)$, or, to put it crudely, no sedimentation) it confers a first-order concentration-dependence on the strength $p_y$ that it should not have if $K$ is constant. A much better way to linearize, if needs must, is to make $K$ constant without expanding the Henky strain. The difference between the two will be illustrated graphically shortly and it is considerable.

In the absence of wall effects, eqns 5,6 and A3.1 can be integrated straightforwardly to give [28, 30],

$$\frac{H_{eq}}{H_0} \cong K^* \ln(1+1/K^*) \equiv \frac{\lambda}{H_0} \ln\left(1 + \frac{H_0}{\lambda}\right) \qquad (A3.2$$

where $\lambda = K^* H_0$ is a length-scale parameter introduced by Manley et al. [28] and used by Condre et al. [30]. Notice that eqn A3.2 can be expanded to give,

$$\frac{H_{eq}}{H_0} \simeq \frac{H_0}{2\lambda} \equiv \frac{1}{2K^*}. \qquad (A3.2a$$

The predictions of A3.2 are compared to exact results calculated for various cases, including constant $K$, i.e. $n = 0$ in eqn 7 and then various powers of $n$ from unity upwards. When the initial gel has strength, there is a critical value of $H_0$ below which the gel is stable. In fig. 7 the initial height $H_0$ is scaled on this where possible. This is not possible for the line corresponding to eqn A3.2 because of the lack of strength at $\varphi_0$ and so this has been positioned arbitrarily (by matching with the curve for $\varphi_0/\varphi_g =$ 2 and $n=1$ at one point such that the the $K^*$ values are equal). It can be seen why it might be possible to fit eqn A3.2 to data for gels near their gel points, since, when it used at finite strains it behaves, in effect, as if there was a concentration dependence of $K$ and $p_y$ that is rather pronounced. It looks nothing at all like the exact results for $n=0$ and $n=1$. Also shown by the three grey curves are exact results very close to the gel-point at $\varphi_0/\varphi_g =1.05$ (where there is still some strength and so $H_c$ has a finite value and can be defined). It can be seen even though these curves would turn over to slope $1/n$ eventually, their initial behaviour is virtually $n$ independent and very like curve A, eqn A3.2. This comparison makes it very clear indeed why it might be possible to fit A3.2 to experimental data fairly near the gel point and, more crucially, near the critical height.

It is possible using eqn A3.1 to derive a linearized analytical result for the full time-dependent sedimentation curve at any $H_0$ [28] and this has likewise been used [28,30] to fit data obtained for low-density gels over a limited range of parameter space. Similar comments apply though. The use such results cannot be justified, except for very small $\Delta = H_o - H(t)$, and, again, any apparent ability to fit data for non-negligible $\Delta = H_o - H(t)$ will be a consequence of the spurious concentration dependence of the strength built in by using A3.1 outside of its range of validity. That this is so can again be demonstrated by means of comparisons with numerical results, although such comparisons will not be presented here, since the point has been made, arguably, given that it is the failure of eqn A3.1 that causes the problems.

In their study of very tenuous gels ( $0.0025 \leq \varphi_0 \leq 0.01$ ) made from nanosize silica particles, Condre et al. [30] found that the equilibrium height scaled like $H_0$ and not like $H_0^2$ as would be expected from A3.2a (for small $\Delta$ only!). They interpreted this, correctly, to mean that the gels had little compressive strength and were supported at equilibrium by static wall friction (i.e. adhesion). That this must be so can be seen readily by analogy to plug flow in a cylindrical tube – the wall stress will support a certain pressure drop, or self-weight in this case, per unit length of tube. Thus, if wall friction dominates, then $H_{eq}/H_0$ becomes independent of $H_0$. Condre et al., derived a linearized eqn for sedimentation equilibrium more generally, analogous to eqn. A3.2, their eqn 9 turns out to incorrect, the correct result is identical to A.3.2 but with the length-scale parameter replaced by a new composite length-scale $\xi$ defined as shown,

$$\frac{H_{eq}}{H_0} \cong \frac{\xi}{H_0}\ln\left(1+\frac{H_0}{\xi}\right); \quad \frac{1}{\xi} = \frac{1}{\lambda} - \frac{1}{L} \tag{A3.3}$$

note that $\xi$ is equal to minus *l*, where *l* is the composite length-scale, introduced by Condre et al. in their erroneous eqn 9. The second new length-scale in eqn A3.3 is given in the notation used here (which is different to that of Condre et al. for reasons of historic usage) by,

$$L = \frac{D}{4S} \tag{A3.4}$$

where, as before, *S* is the ratio of wall shear strength to compressive strength. Notice then that A3.3 reduces to A3.2 in the limits of large diameter, small shear strength and large compressive strength, as it should.

Condre et al. [30] were caused by their data to suppose that the wall stress (or static wall friction) was dependent upon the radial stress in the network and hence, in the terms used in this paper, upon $p_y$. They interpreted this apparent pressure-dependence to mean that the wall-friction was coloumbic, and embarked upon a lengthy discussion of this point, which, it has to be said, the author of this paper finds unconvincing, why would such tenuous gels behave like a sandy soil? And, are not these gels sticky and thus adhesive intrinsically? Be that as it may, it turns out to not to be *necessary* to invoke granular friction since the pressure-dependence need only be apparent and not causal as will now be shown. The critical quantity here is the ratio of shear to compressive strength appearing in the definition of the new length-scale *L*. The compressive strength $p_y$ is concentration-dependent of course, even in the linearized model and it would be strange indeed if the adhesive strength were not also – gels become stronger as they are concentrated. Notice now that from eqns 2 and 3, a shared dependence of the shear and compressive strengths on volume-fraction leads to an *apparent* dependence of the wall stress on the pressure, this is an inevitable consequence of eqn 3 in particular, which states that the network pressure obeys $p_s \leq p_y(\varphi)$. So, whereas it cannot be said categorically that Condre et al. are wrong to invoke granular friction and also the possibility of anisotropy and, perhaps, cross-elasticity (via their redirection coefficient), there is certainly no need to invoke it. Adhesion independent of pressure suffices, provided that one makes the very reasonable assumption that the wall stress increases with concentration.

The remaining point to address is the question of why Condre et al [30] encountered a friction-dominated regimen in their experiments when it is more usual (*vide infra*) to see compressional effects dominating. Although the rheological data needed to give a hard and fast quantitative explanation is lacking, three features of the gels used by Condre would seem to be relevant here in the context of the model proposed in the form of eqn 17 and bearing mind that the ratio $D/4H_0$ is important also via eqns A3.3 and A3.4. These features are,

i) Very low density: 0.0025-0.01 (and thus near the gel point).

ii) Large elastic strain limit in shear (reported as ca. 0.1[30], with this in turn being as result of the combination of low density and very small particle size, it is surmised).

iii) $D/4H_0 < 1$.

It is interesting to note then, in the context of point (i) that Manley et al. [28], who used more concentrated but otherwise very similar gels, saw compressive effects as more important. The acid test though would be a more extended sweep across the volume-fraction range, using centrifuges if needs be.

**Figures follow, one to a page, seven in all.**

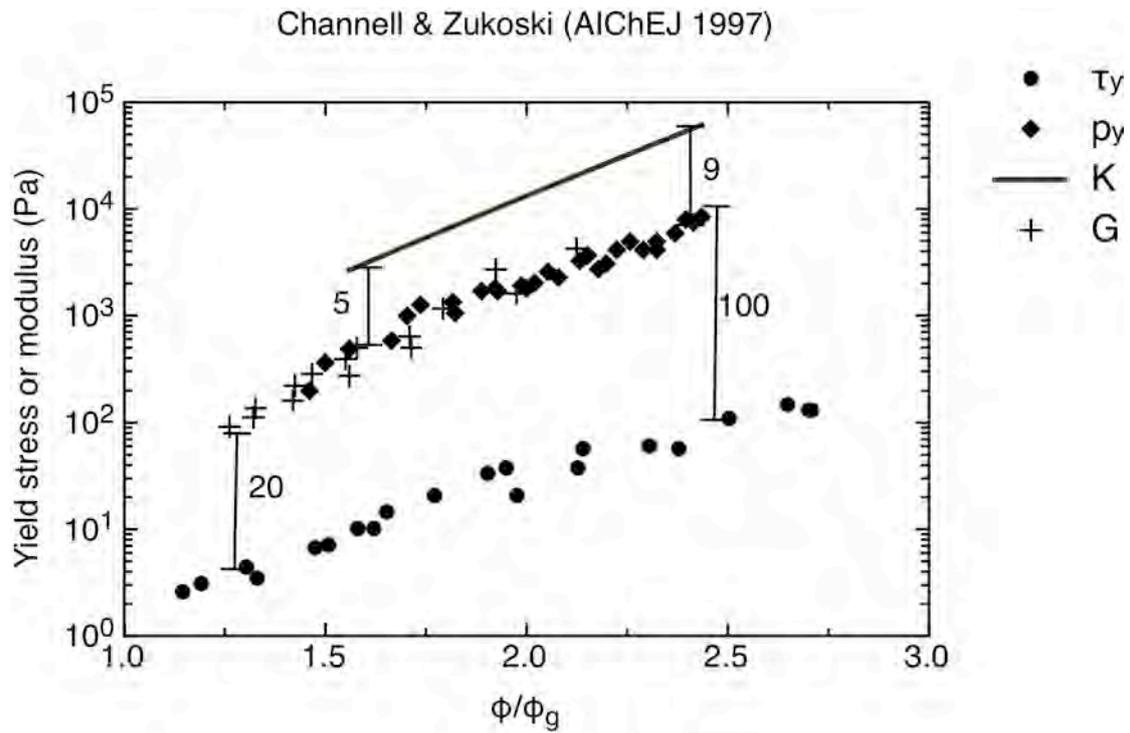

**Fig. 1** Plot of shear strength, $\tau_y$, compressional strength, $p_y$, shear modulus, $G$, and compressional modulus, $K$, versus volume-fraction, normalised by that at the gel-point, for flocculated α-alumina particles – data taken from Channell & Zukoski [38]. It should be noted that the $K$ values differ somewhat from those shown in [38] which are erroneous.

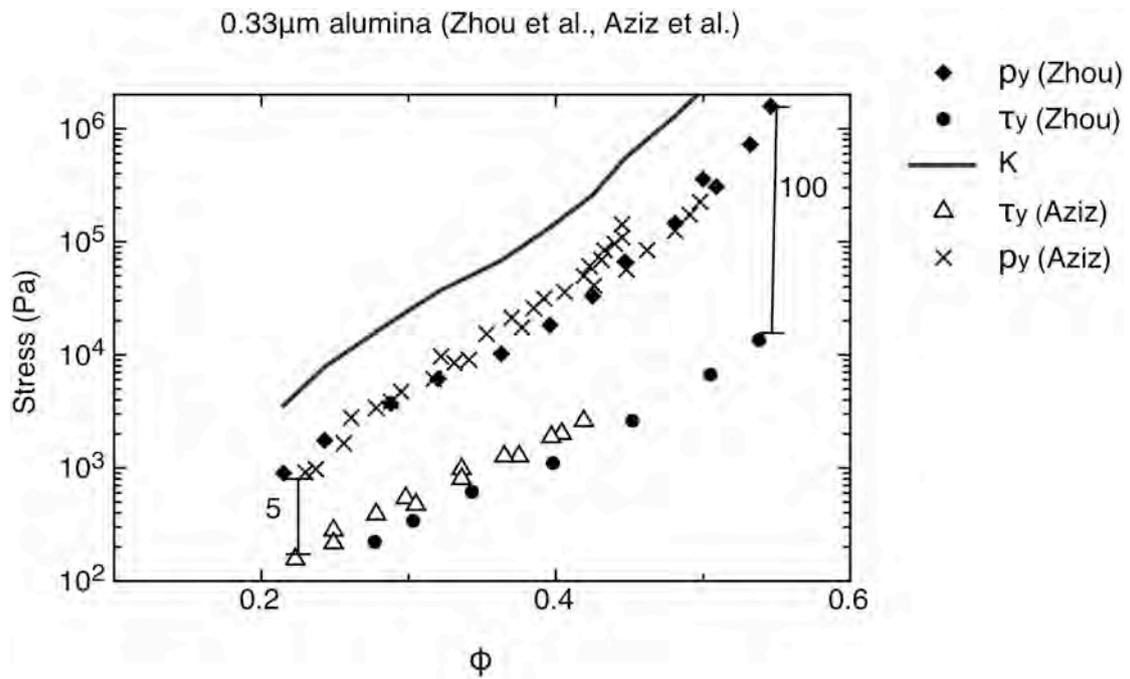

**Fig. 2** Plot of shear strength, $\tau_y$, compressional strength, $p_y$, and compressional modulus, $K$, versus volume-fraction, normalised by that at the gel-point, for flocculated α-alumina particles. Data from the Melbourne group [42-44]. Note the good agreement between two separate investigations.

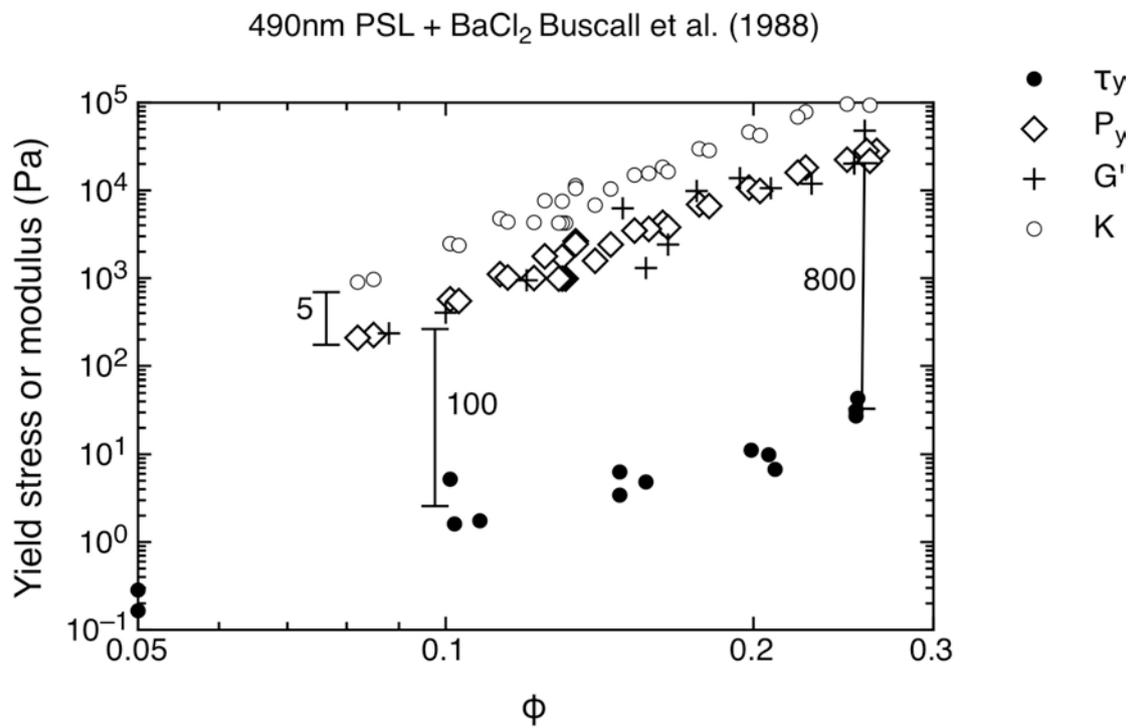

**Fig. 3** Plot of shear strength, $\tau_y$, compressional strength, $p_y$, shear modulus, $G$, and compressional modulus, $K$, versus volume-fraction for flocculated PSL particles – data taken from Buscall et al. [37, 40-41].

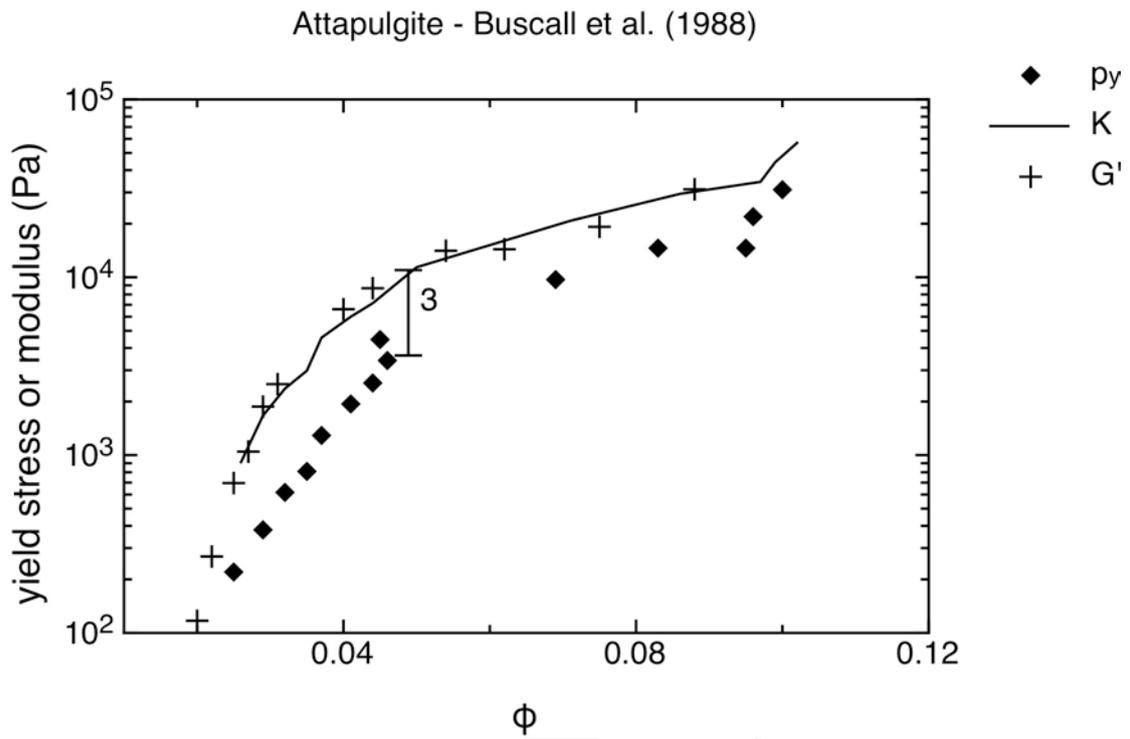

**Fig. 4** Plot of compressional strength, $p_y$, shear modulus, $G$, and compressional modulus, $K$, versus volume-fraction for flocculated attapulgite clay particles – data taken from Buscall et al. [41].

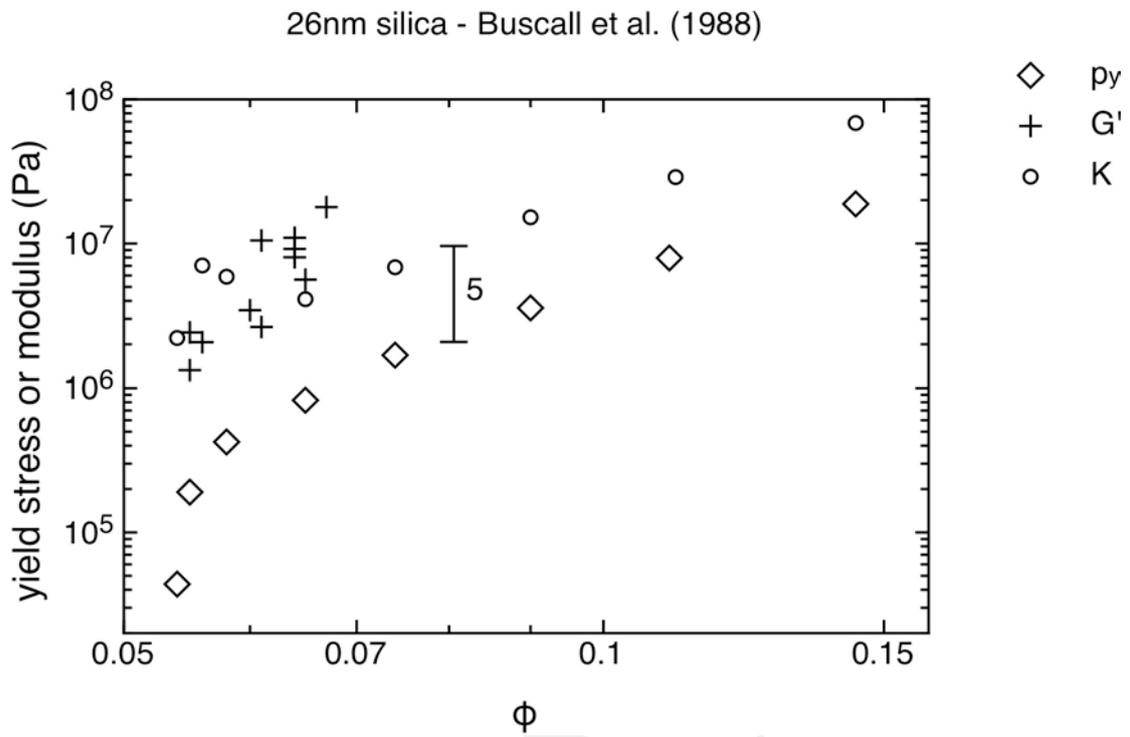

**Fig. 5** Plot of compressional strength, $p_y$, shear modulus, $G$, and compressional modulus, $K$, versus volume-fraction for flocculated silica particles – data taken from Buscall et al. [41].

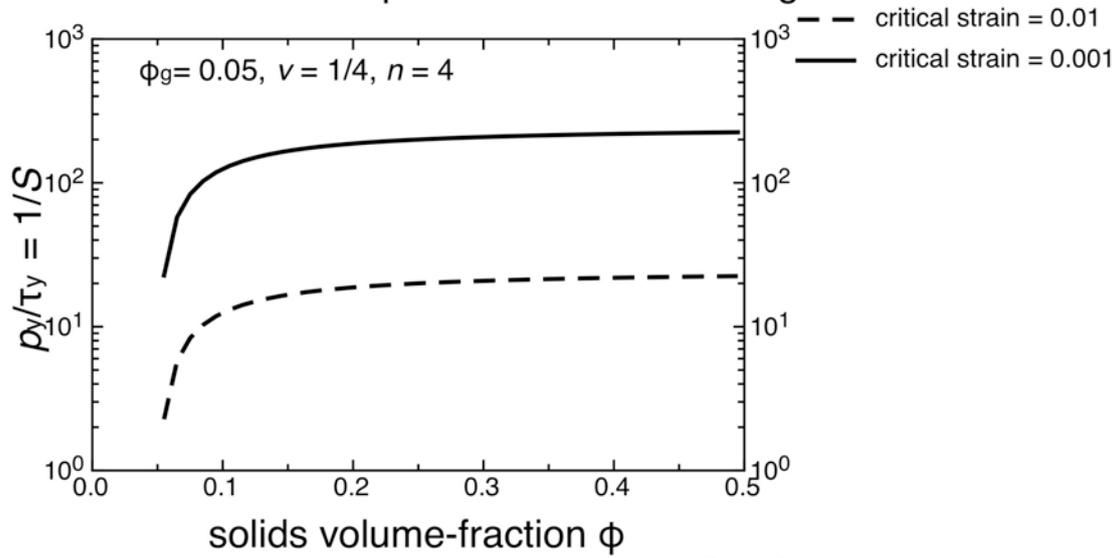

**Fig. 6** Some predicted values for the strength ratio for two values of the critical shear strain. Otherwise, the gel-point, the poisson's rati and the concentrationpower-law index have the values shown. Channell and Zukoski [38], who measured critical shear strains of order 0.01, saw values of $1/S$ ranging from of ca. 20 at ca. 1.25 $\varphi_g$ to ca. 100 100 at high volume-fraction (see fig. 1).

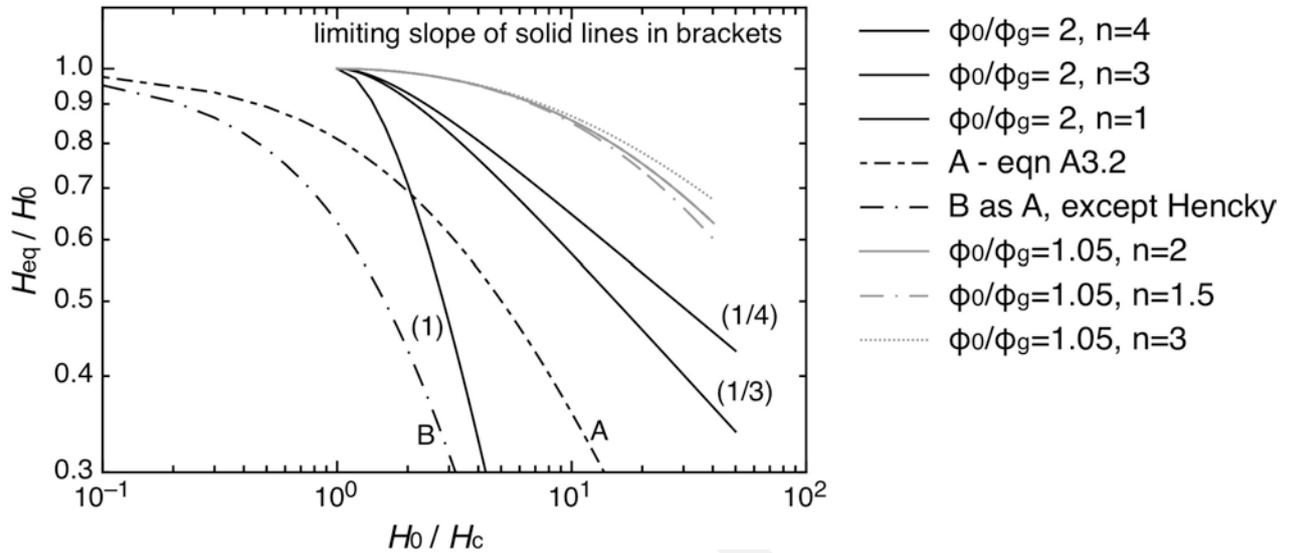

**Fig. 7** Theoretical values of dimensionless equilibrium sediment height versus initial height scaled on critical height. The solid black lines show exact results (eqns 8) at twice the gel-point and for $n = 1$, 3 and 4. Note that the limiting slopes are equal to $1/n$. Curve A is the fully linearised approximation, whereas as curve B is constant $K$ but full Hencky strain, but with the initial strength set to zero so as to compare with A otherwise. Finally, the three grey curves are exact results very close to the gel-point.